\documentclass[namedreferences]{solarphysics}
\usepackage[hyperref,optionalrh]{spr-sola-addons} 
\usepackage{graphicx}        
\usepackage{color}           
\usepackage{breakurl}        
\def\degree{${}^{\circ}$}

\newcommand{\aap}{    {\it Astron. Astrophys.}}

\newcommand{\apj}{    {\it Astrophys. J.}}
\newcommand{\apjl}{   {\it Astrophys. J. Lett.}}
\newcommand{\apjs}{   {\it Astrophys. J. Suppl.}}

\newcommand{\jgr}{    {\it J. Geophys. Res.}}

\newcommand{\prl}{   {\it Phys. Rev. Lett.}}

\newcommand{\solphys}{{\it Solar Phys.}}

\newcommand{\ssr}{    {\it Space Sci. Rev.}}
\chardef\us=`\_
\begin{document}

\begin{article}
\begin{opening}

\title{Formation and Immediate Deformation of a Small Filament Through Intermittent Magnetic Interactions}

\author[addressref={aff1,aff2},corref,email={ruishengzheng@sdu.edu.cn}]{\inits{R.}\fnm{Ruisheng}~\lnm{Zheng}\orcid{0000-0002-2734-8969}}
\author[addressref={aff1},email={zhangliang1121@mail.sdu.edu.cn}]{\inits{L.}\fnm{Liang}~\lnm{Zhang}\orcid{0000-0001-6205-2496}}
\author[addressref={aff3,aff4},corref,email={chrao@ioe.ac.cn}]{\inits{C.}\fnm{Changhui}~\lnm{Rao}}
\author[addressref={aff1},email={wbing@sdu.edu.cn}]{\inits{B.}\fnm{Bing}~\lnm{Wang}}
\author[addressref={aff2,aff5},email={hdchen@nao.cas.cn}]{\inits{H.}\fnm{Huadong}~\lnm{Chen}\orcid{0000-0001-6076-9370}}
\author[addressref={aff3,aff4},email={zhonglibo@ioe.ac.cn}]{\inits{L.}\fnm{Libo}~\lnm{Zhong}}
\author[addressref=aff1,email={yaochen@sdu.edu.cn}]{\inits{Y.}\fnm{Yao}~\lnm{Chen}\orcid{0000-0001-6449-8838}}

\address[id=aff1]{Shandong Key Laboratory of Optical Astronomy and Solar-Terrestrial Environment, School of Space Science and Physics, Institute of Space Sciences, Institute of Frontier and Interdisciplinary Science, Shandong University, Weihai, Shandong, 264209, China}
\address[id=aff2]{CAS Key Laboratory of Solar Activity, National Astronomical Observatories, Chinese Academy of Sciences, Beijing 100101, China}
\address[id=aff3]{The Key Laboratory on Adaptive Optics, Chinese Academy of Sciences, P.O. Box 350, Shuangliu, Chengdu 610209, Sichuan, China}
\address[id=aff4]{The Laboratory on Adaptive Optics, Institute of Optics and Electronics, Chinese Academy of Sciences, P.O. Box 350, Shuangliu, Chengdu 610209, Sichuan, China}
\address[id=aff5]{School of Astronomy and Space Science, University of Chinese Academy of Sciences, Beijing 100049, China}

\runningauthor{L. Zhang et al.}
\runningtitle{Formation and Immediate Deformation of a Small Filament}

\begin{abstract}
It is generally believed that filament formation involves a process of the accumulation of magnetic energy. However, in this paper we discuss the idea that filaments will not erupt and will only deform when the stored magnetic energy is released gradually. Combining high-quality observations from S\emph{olar} D\emph{ynamics} O\emph{bservatory} and other instruments, we present the formation and immediate deformation of a small filament (F1) in the active region (AR) 12760 on 28\,--\,30 April 2020. Before the filament formation, three successive dipoles quickly emerged with separation motions in the center of AR 12760. Due to the magnetic interaction between magnetic dipoles and pre-existing positive polarities, coronal brightenings consequently appeared in the overlying atmosphere.
Subsequently, because of the continuous cancellation of magnetic flux that happened around the adjacent ends of F1 and another nearby filament (F2), the magnetic reconections occurred intermittently occurred between F1 and F2. Finally, F1 lessened in the shear, and F2 became shorter. All the results show that the formation of F1 was closely associated with intermittent interactions between the sequence of emerging dipoles and pre-existing magnetic polarities, and the immediate deformation of F1 was intimately related to intermittent interactions between F1 and F2. We also suggest that the intermittent magnetic interactions driven by the continuous magnetic activities (magnetic-flux emergence, cancellation, and convergence) play an important role in the formation and deformation of filaments.
\end{abstract}
\keywords{Sun: activity --- Sun: corona --- Sun: magnetic fields --- Sun: filaments, prominences}
\end{opening}
\section{Introduction}
     \label{S-Introduction}
Solar filaments consist of cold, and dense plasma suspended in the hot corona, and are always located along polarity inversion lines (PILs) separating opposite polarities of photospheric radial fields~\citep{1998Martin, 2010Mackay}. A common feature in the solar corona, filaments are an active subject of research in solar physics because a filament eruption can release a large amount of plasma and magnetic field and thus lead to a significant disturbance for space weather. The magnetic energy released in the eruption process is previously stored in highly stressed/twisted coronal structures~\citep{2000Forbes, 2001Low}. Hence, filament formation is the process of the accumulation of magnetic energy.

The formation mechanism of filaments is one main issue under debate. On the one hand, filaments are thought to be formed in the solar interior and emerged into the atmosphere through magnetic buoyancy~\citep{1994Low, 2001Fan, 2013Leake}.~\cite{2009fan} studied a twisted flux rope that emerged from the solar interior to the corona through magnetohydrodynamic (MHD) simulation, and~\cite{2017yan} presented the emergence of a small-scale flux rope combing with the observations from N\emph{ew} S\emph{olar} T\emph{elescope} and S\emph{olar} D\emph{ynamics} O\emph{bservatory} (SDO). On the other hand, some studies have suggested that filaments are directly formed in the corona through magnetic reconnections, which could be driven by surface motions such as shear flows along the PILs, converging flows to PILs, differential rotation, and magnetic cancellation~\citep{1989van, 2016Yang, 2018Threlfall}.
Some observations revealed that the surface motion played important roles in the filament formation.~\cite{2001Chae} reported that the formation of a filament was closely related to a large cancelling magnetic feature.~\cite{2015Yan} found that two successive filaments were formed by shear flows between opposite polarity magnetic fields and the rotation of a small sunspot. Recently,~\cite{2020Zheng} presented an example of a tiny flux rope formed by magnetic-flux emergence, cancellation, and convergence.

After formation, the magnetic energy stored in the filament structure can be intensely released in an eruption process. It is believed that filament eruptions are intimately related with solar flares and coronal mass ejections (CMEs), and these three eruptive phenomena are different parts of a solar eruption~\citep{2001Zhang}. Many theoretical models have been proposed to interpret the trigger mechanisms of solar eruptions,
such as the interior reconnection tether-cutting model~\citep{1992Moore, 2001Moore}, the external reconnection breakout model~\citep{1999Antiochos}, the magnetic-flux cancellation or emergence model~\citep{1977Heyvaerts, 2000Chen}, the MHD helical-kink instability model~\citep{2004T, 2005Fan}, and the MHD torus instability model~\citep{1978Bateman, 2006Kliem}. In addition, filament eruptions occasionally are fully failed/confined~\citep{2003Ji, 2009Liu, 2010Guo,2013Joshi, 2014Song, 2014Dalmasse, 2017Zhenga, 2018Yang} or partially successful~\citep{2007Gilbert, 2007Liu, 2012Shen, 2015Zhang, 2017Zhengb, 2018Cheng}.

On the other hand, the stored magnetic energy can also be gently released in a few times, and then the filament will only experience deformation, not eruption. In this work, we present the formation and immediate deformation of a small filament in active region (AR) center, and both formation and deformation were closely associated with intermittent magnetic interactions involving magnetic-flux emergence, cancellation, and convergence.
\section{Observations and Data Analysis} 
      \label{S-general}
The formation and deformation of a small filament that occurred in NOAA AR 12760 ($\approx$S06W29) during the period from 28 to 30 April 2020.
The filament was captured by H$\alpha$ filtergrams from the S\emph{olar} M\emph{agnetic} A\emph{ctivity} R\emph{esearch} T\emph{elescope}~\citep[SMART:][]{2004UeNo} with the solar imaging system of the S\emph{olar} D\emph{ynamics} D\emph{oppler} I\emph{mager}~\citep{2017Ichimoto} at the Hida Observatory, and the H$\alpha$ images have a cadence of one minute and a pixel size of $\approx$1$^{\prime \prime}$. Furthermore, some high-resolution H$\alpha$ filtergrams from the 1.8-meter C\emph{hinese} L\emph{arge} S\emph{olar} T\emph{elescope}~\citep[CLST:][]{2015Rao,2020Rao} are employed as a supplement, with a pixel size of 0.077$^{\prime \prime}$.

In addition, we employed full-disk observations from the H\emph{elioseismic} and M\emph{agnetic} I\emph{mager}~\citep[HMI:][]{2012Scherrer} and from the A\emph{tmospheric} I\emph{maging} A\emph{ssembly}~\citep[AIA:][]{2012Lemen} onboard the SDO~\citep{2012Pesnell}. The magnetic-field evolution of the source region during the formation process and the subsequent magnetic cancellation beneath two adjacent ends of filaments was checked by magnetograms and intensity maps from the HMI, with a cadence of 45 seconds and pixel size of 0.6$^{\prime \prime}$. The coronal response was recorded by the AIA images in extreme ultraviolet (EUV) wavelengths, with a pixel size of 0.6$^{\prime \prime}$ and a cadence of 12 seconds. The deformation of the filament was also scanned in narrowband slit-jaw images (SJIs) from the I\emph{nterface} R\emph{egion} I\emph{maging} S\emph{pectrograph}~\citep[IRIS:][]{2014De}, with a cadence of 36 seconds and a pixel size of 0.332$^{\prime \prime}$.

The observations from the different instruments were co-aligned by the obvious identical features (e.g. filaments and brightenings) by eye in the temporally closest images. Several objects were  co-aligned simultaneously and thus, to a great extent, ensure that the cross-data comparisons we make are accurate and reliable (approximately good to 5$^{\prime \prime}$).

\section{Results} 
      \label{S-general}
\subsection{Filament Formation}
\begin{figure}
	\includegraphics[width=\columnwidth]{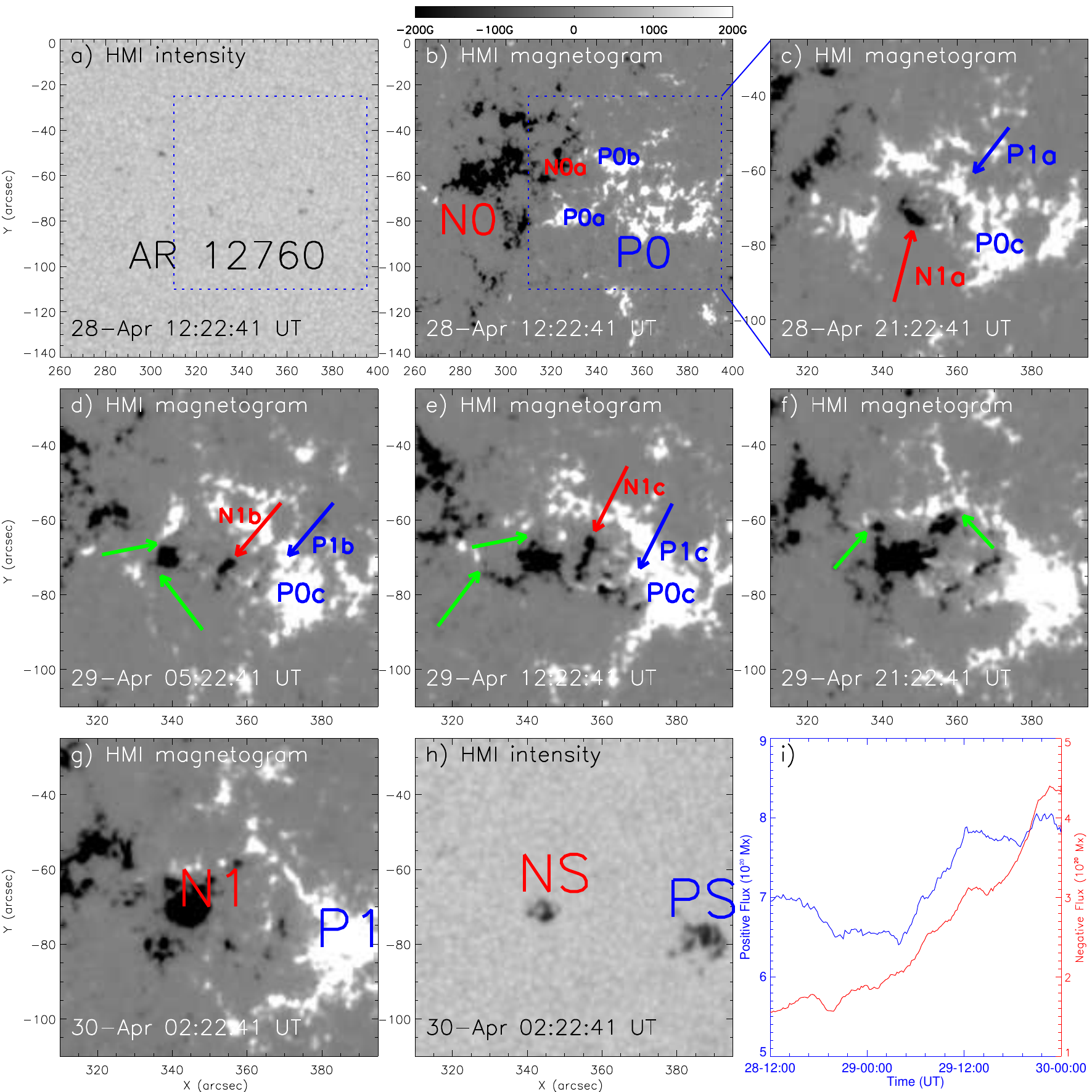}
    \caption{Overview of AR 12760 in HMI magnetogram and HMI intensity map (\emph{\textbf{a\,--\,b}}). The main polarities N0-P0 and relative small polarities N0a, P0a, P0b. The \emph{blue-dashed boxes} indicate the field view of \emph{\textbf{c\,--\,h}}. The magnetic-field evolution in the \emph{region of dashed box} in the sequence of HMI magnetogram maps \emph{\textbf{c\,--\,g}} and HMI intensity map \emph{\textbf{h}} during the formation process. The changes of positive and negative flux for this region \emph{\textbf{i}}. The \emph{blue and red arrows} indicate the emerged positive and negative polarities, respectively. The \emph{green arrows} indicate the interaction sites. The \emph{blue and red lines} indicate the positive and negative flux, respectively. An accompanying animation is available in the Electronic Supplementary Material. The \emph{left and right panel} of the animation show the evolution of the HMI intensity and magnetogram map, respectively. The animation begins at 12:22:41 UT on 28 April and ends at 00:22:41 UT on 30 April. The cadence is 19 seconds.}
    \label{fig1}
\end{figure}

Figure~\ref{fig1} shows the magnetic-field evolution of the source region (dashed boxes) in AR 12760. Initially, AR 12760 only appeared as faint tiny pores in HMI-intensity images (Panel a), and simply consisted of a leading positive polarity (P0) and a following negative polarity (N0) in HMI magnetograms (Panel b). At the interface between P0 and N0, some minor polarities (N0a and P0a-P0b) were important for the following filament formation (Panel b). After 12:00 UT 28 April, some dipoles (P1a-N1a, P1b-N1b, P1c-N1c) successively emerged from the gap between P0a-P0b and the majority (P0c) of P0, and the negative and positive polarities (red and blue arrows) moved and converged eastwards and westwards, respectively (Panels c\,--\,f). As a result of the continuous emergence and convergence, the compact positive (P1) and negative (N1) polarities finally formed at the beginning of 30 April, corresponding to the distinct positive and negative sunspots (PS and NS), respectively (Panels g\,--\,h). The emergence of dipoles is consistent with the obvious enhancement of positive and negative magnetic flux in the source region (Panel i).
Besides, because of the separation of dipoles, the eastward negative polarities (N1a, N1b, N1c) began to interact with pre-existing P0a-P0b (green arrows in Panels d\,--\,f), which resulted in the positive polarities nearly vanishing between N0 and newly formed N1 (Panel g).

\begin{figure}
	\includegraphics[width=\columnwidth]{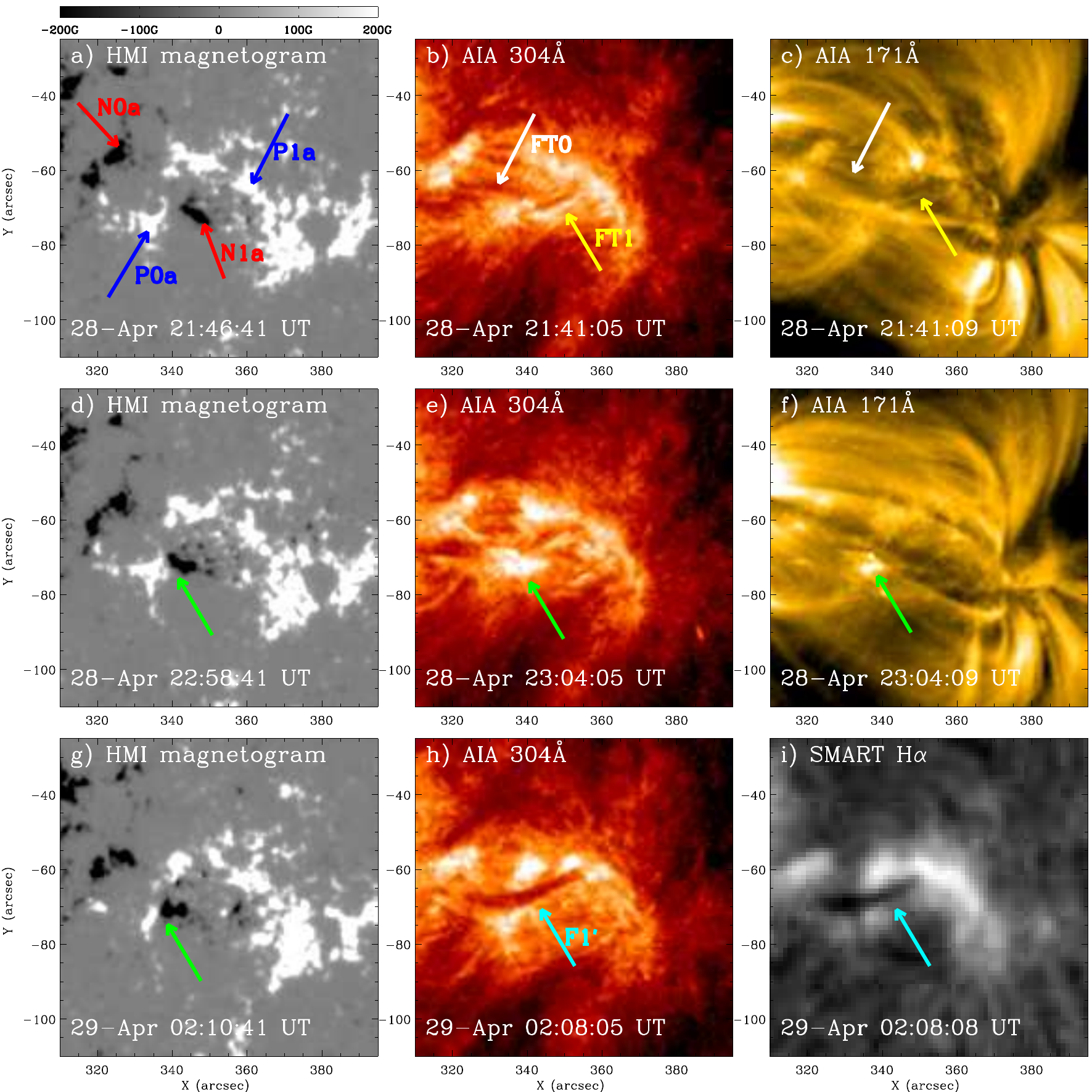}
    \caption{The first magnetic interaction associated with the dipole of P1a-N1a and the formation of the first-phase concave filament F1$^{\prime}$ in HMI magnetograms (\emph{left column}), AIA 304{\AA} (\emph{middle column}), 171{\AA}  \emph{\textbf{c}} and \emph{\textbf{f}}, and SMART H$\alpha$ filtergram \emph{\textbf{i}}. The emerged polarities N1a-P1a and pre-existing polarities N0a, P0b (\emph{red and blue arrows}). The \emph{white and yellow arrows} indicate the filament threads FT0 and FT1, respectively. The \emph{green arrows} indicate the cancellation site and associated brightening in the upper atmosphere. The \emph{cyan arrows} indicate the first-phase concave filament F1$^{\prime}$.}
    \label{fig2}
\end{figure}

The formation of a small filament in the center of AR 12760 was closely associated with intermittent magnetic interactions between three emerging dipoles and the pre-existing magnetic polarities (Figures 2\,--\,4). Figure~\ref{fig2} illustrates the first magnetic interaction associated with the P1a-N1a dipole. Some filament threads (FT0; white arrows) rooted at P0a and N0a before the emergence of P1a-N1a that brought the appearance of the filament threads (FT1; blue arrows) connecting P1a and N1a (upper panels). During the emergence, N1a moved eastwards close to P0a. Due to the approach, some brightenings appeared clearly in the above atmosphere (green arrows in Panels e\,--\,f), which is the evidence of magnetic cancellation between N1a and P0a (green arrows in Panels d and g). As a result, a small concave (i.e. the center of the filament is curved towards the center of the active region) filament (F1$^{\prime}$) formed connecting N0a and P1a (cyan arrows in bottom panels).

\begin{figure}
	\includegraphics[width=\columnwidth]{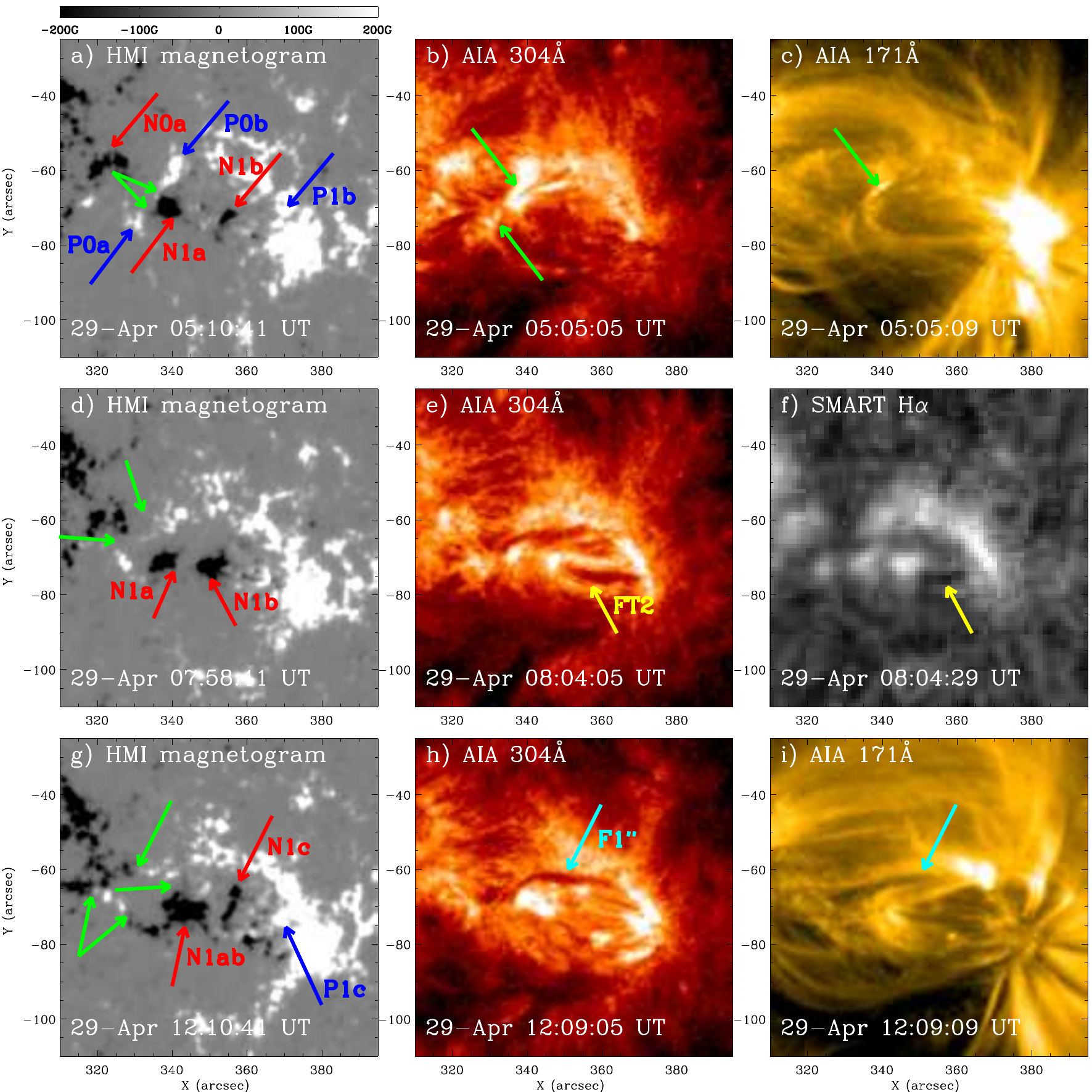}
    \caption{The second magnetic interaction associated with the dipole of P1b-N1b and the formation of the second-phase convex filament F1$^{\prime\prime}$ in HMI magnetogram map (\emph{left column}), AIA 304{\AA} (\emph{middle column}), 171{\AA} \emph{\textbf{c}} and \emph{\textbf{i}}, and SMART H$\alpha$ filtergram \emph{\textbf{f}}. The newly emerged polarities N1b-P1b, N1c-P1c, the pre-existing polarities N0a, P0a, P0b, N1a and the merged polarities N1ab (\emph{red and blue arrows}). The \emph{green arrows} indicate the cancellation site and associated brightening in the upper atmosphere. The \emph{yellow arrows} indicate the filament threads FT2. The \emph{cyan arrows} indicate the second-phase convex filament F1$^{\prime\prime}$}.
    \label{fig3}
\end{figure}

The second magnetic interaction associated with the dipole of P1b-N1b is shown in Figure~\ref{fig3}. During the emergence of P1b-N1b, N1a not only kept interacting with P0a, but also moved northwards and interacted with P0b (green arrows in Panel a). As a result, some brightenings appeared in AIA 304 and 171{\AA} (green arrows in Panels b\,--\,c). Since the emergence, N1b also moved eastwards and mixed with N1a (red arrows in Panels d and g). The stronger N1ab continued to interact with both P0a and P0b, and at the same time, N0a also canceled with P0a and P0b (green arrows in Panel g). As a result, P0a and P0b became fainter and fainter (left panels). Besides, the overlying concave F1$^{\prime}$ became convex (i.e. the center of the filament is curved away from the center of the active region) F1$^{\prime\prime}$ (cyan arrows in Panels h\,--\,i). Note that the newly emerging filament threads (FT2; yellow arrows in Panels e\,--\,f) connecting P1b and N1b barely contributed to the filament formation. During the second interaction, the third dipole of P1c-N1c quickly emerged (Panel g).

\begin{figure}
	\includegraphics[width=\columnwidth]{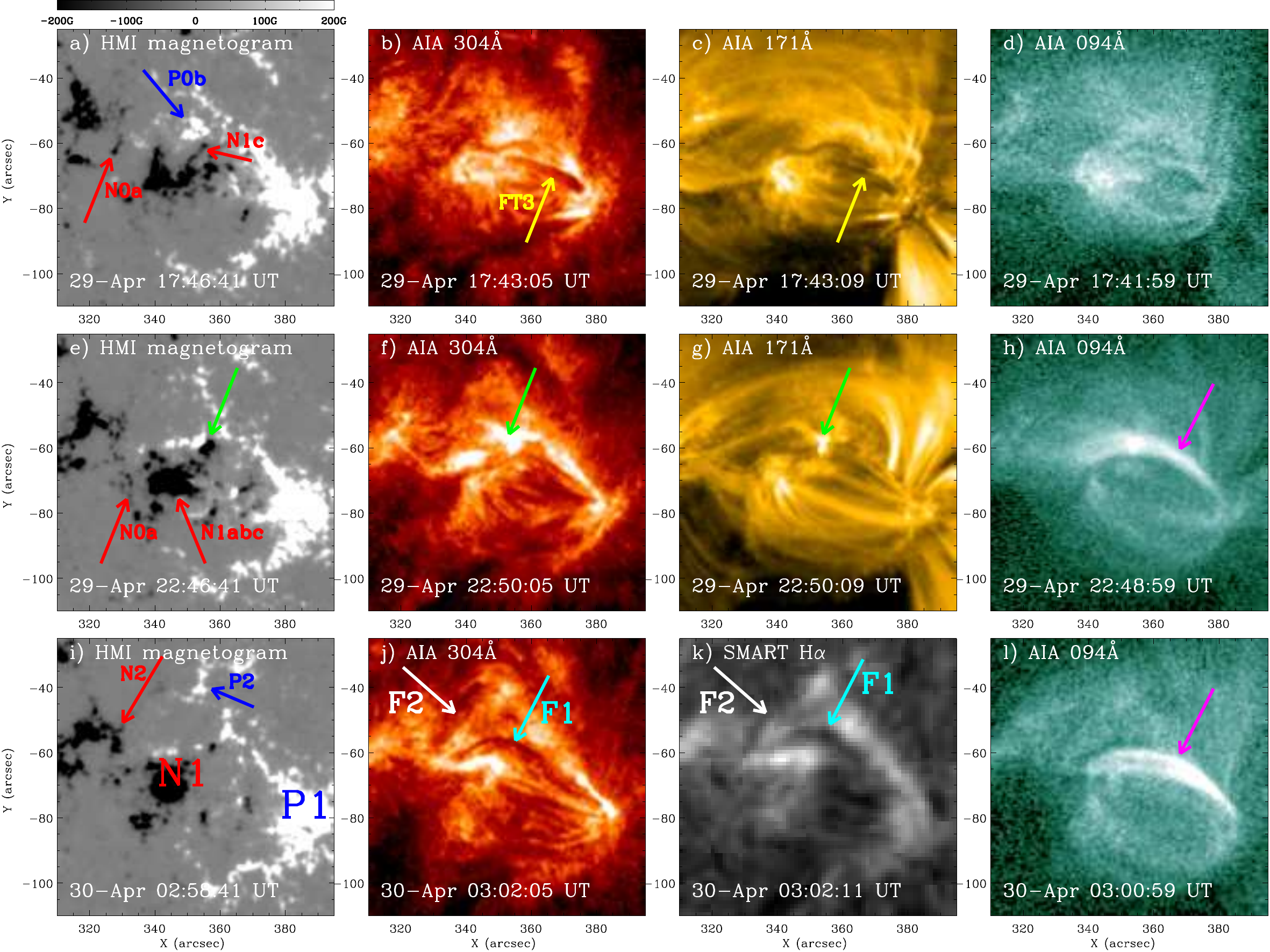}
    \caption{The third magnetic interaction associated with the dipole of P1c-N1c and the formation of the elongated filament F1 in HMI magnetogram map (\emph{left column}), AIA 304{\AA} (\emph{second left column}), 171{\AA} \emph{\textbf{c, g}}, 94{\AA} (\emph{right column}) and SMART H$\alpha$ filtergram \emph{\textbf{k}}. The polarities N0a, P0b, N1c and the merged polarities N1abc (\emph{red and blue arrows}). The magnetic-field of the source region after F1 formed and polarities N1-P1 and N2-P2 \emph{\textbf{i}}. The \emph{yellow arrows} indicate the filament threads FT3. The \emph{green arrows} indicate the cancellation site and associated brightening in the upper atmosphere. The \emph{cyan and pink arrows} indicate the newly formed filament F1 and its overlying hot loops, respectively. The \emph{white arrows} indicate another short filament F2 located in the northeast to F1.}
    \label{fig4}
\end{figure}

The third magnetic interaction associated with the dipole of P1c-N1c is shown in Figure~\ref{fig4}. During the emergence and separation, the filament threads (FT3; yellow arrows) connecting P1c and N1c became longer, and N1c moved towards P0b (upper panels). The approach lead to the magnetic interaction between N1c and P0b and the following brightenings in the atmosphere above (green arrows in second panels). Besides, N0a moved towards and mixed with N1abc, and it constituted part of compact N1 (Panel i). Finally, an elongated filament (F1; cyan arrows in bottom panels) formed, with two ends rooting at N1 and P1. Overlying F1, a sheared hot loop in AIA 94{\AA} appeared (pink arrows in Panels h and l). Meanwhile, another shorter filament (F2; white arrows) connecting another dipole of P2-N2 located in the northeast of F1, and the south end of F2 was close to the east end of F1 (bottom panels).

\subsection{Filament Deformation}

\begin{figure}
	\includegraphics[width=\columnwidth]{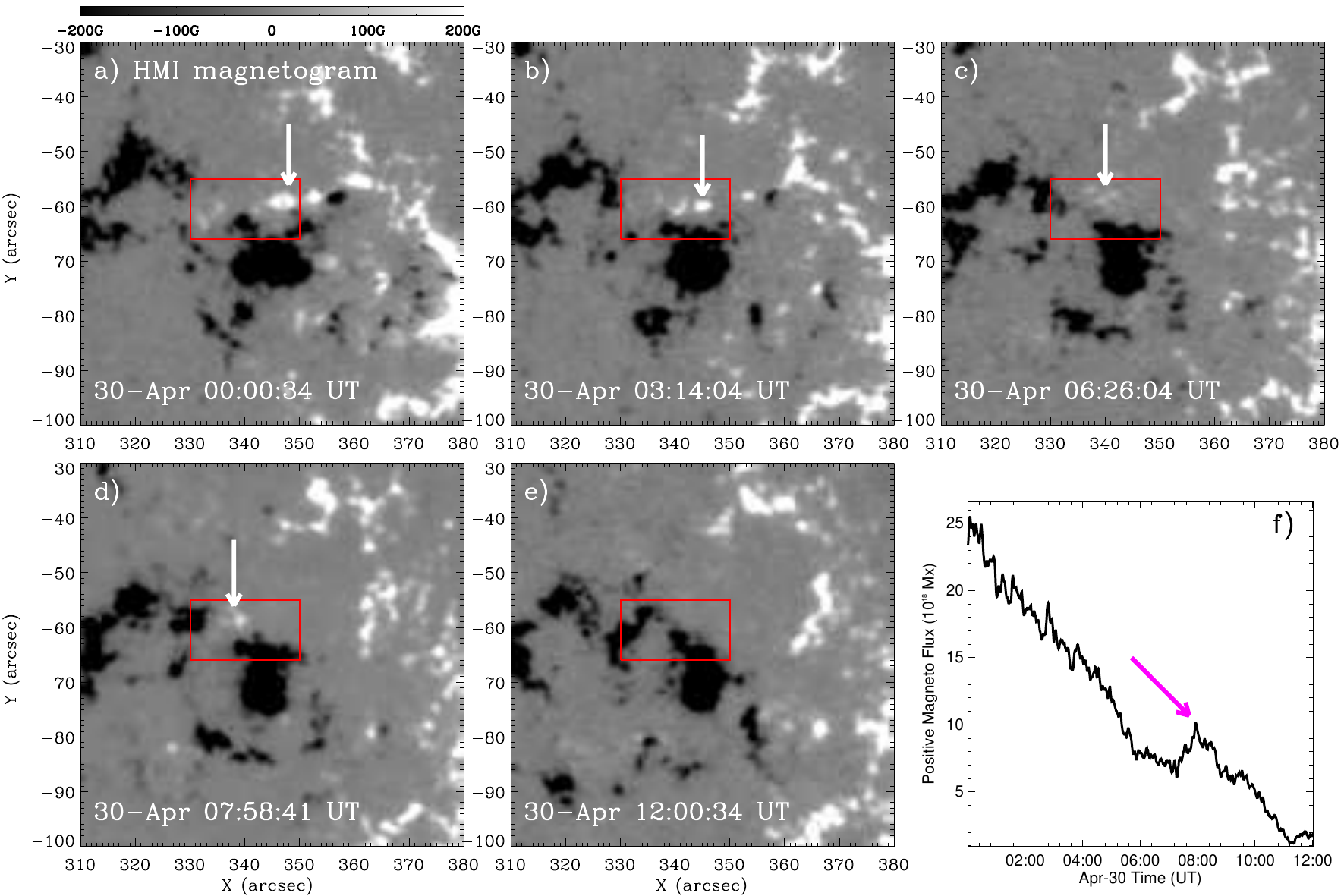}
    \caption{The magnetic-field evolution beneath two filaments in HMI magnetograms \emph{\textbf{a\,--\,e}}. The \emph{red boxes} indicate the region around N1-N2 where some micro positive polarities disappeared and emerged and vanished again, and the changes of positive flux for this region \emph{\textbf{f}}. The \emph{white arrows} indicate the micro positive polarities. The \emph{pink arrow} indicates the bump in the decreasing curve. An accompanying animation is available in the Electronic Supplementary Material. The animation shows the evolution of the HMI magnetogram. The animation begins at 00:00:34 UT and ends at 12:00:34 UT on 30 April. The cadence is 20 seconds.}
    \label{fig5}
\end{figure}

The magnetic-field evolution beneath the two filaments is shown in Figure~\ref{fig5}. In the sequence of HMI magnetograms (Panels a\,--\,e), it is obvious that some micro positive polarities beneath the adjacent negative ends of filaments (N1 and N2) slowly disappeared and emerged and vanished again (white arrows). The magnetic-flux evolution (between 00:00 and 12:00 UT on 30 April) in the region around N1-N2 (red boxes) is shown in Panel f. The positive flux continuously decreased and nearly vanished at 12:00 UT, with a net decrease of $\approx25 \times 10^{18}$ Mx. In the decreasing curve, there a bump at $\approx$08:00 UT (pink arrow), which is consistent with the abrupt emergence and quick disappearance of the minor positive polarity in Panels d\,--\,e. The curve indicates the continuous magnetic-flux cancellation around the negative ends of two filaments. As a result, the north edge of N1 profile was distorted.

\begin{figure}
	\includegraphics[width=\columnwidth]{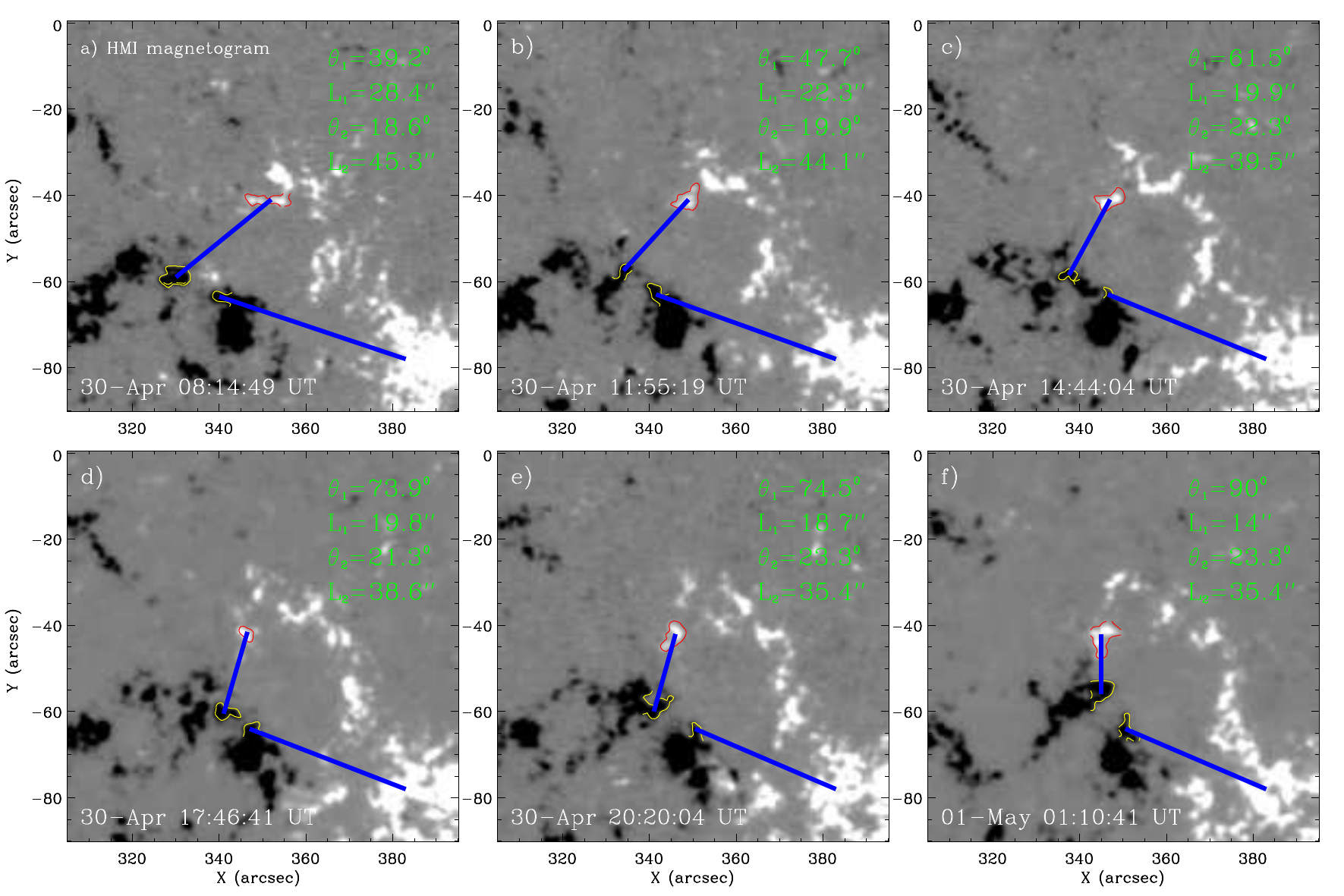}
    \caption{The convergence motion and anticlockwise rotation of the minor polarities N2, P2 related to two ends of F2 in  HMI magnetograms. The \emph{yellow and red contours} indicate the westernmost piece of N2 and the southernmost piece of P2 (N2$^{\prime}$ and P2$^{\prime}$). The \emph{blue lines} connect the centers of N2$^{\prime}$ and P2$^{\prime}$. \emph{L$_1$ and $\theta_1$} indicate the length of blue line and the angle between L$_1$ and the horizontal line. The values are listed on the upper-right corner of each panel.}
    \label{fig6}
\end{figure}

Moreover, as the continuous magnetic-flux cancellation beneath the adjacent filament ends, the minor polarities of P2 and N2 moved towards each other. The convergence of P2-N2 is shown in Figure~\ref{fig6}. We tracked the closest part of two polarities, the westernmost piece of N2 and the southernmost piece of P2 (N2$^{\prime}$ and P2$^{\prime}$; yellow and red contours), and connect the centers of N2$^{\prime}$ and P2$^{\prime}$ with the blue line. The length of the blue line ($L_1$) and the angle $\theta$$_1$ between the blue line and the horizontal line are used to analyse the convergence motion. In the sequence of HMI magnetograms, $L_1$ gradually decreased from $\approx$28.4$^{\prime\prime}$ to $\approx$14$^{\prime\prime}$, and $\theta$$_1$ gradually increased from $\approx$39.2\degree to $\approx$90\degree with a speed of $\approx$2.9\degree h$^{-1}$. It indicates that the minor polarities related to the two ends of F2 experienced a convergence motion, with a counterclockwise rotation.

\begin{figure}
	\includegraphics[width=\columnwidth]{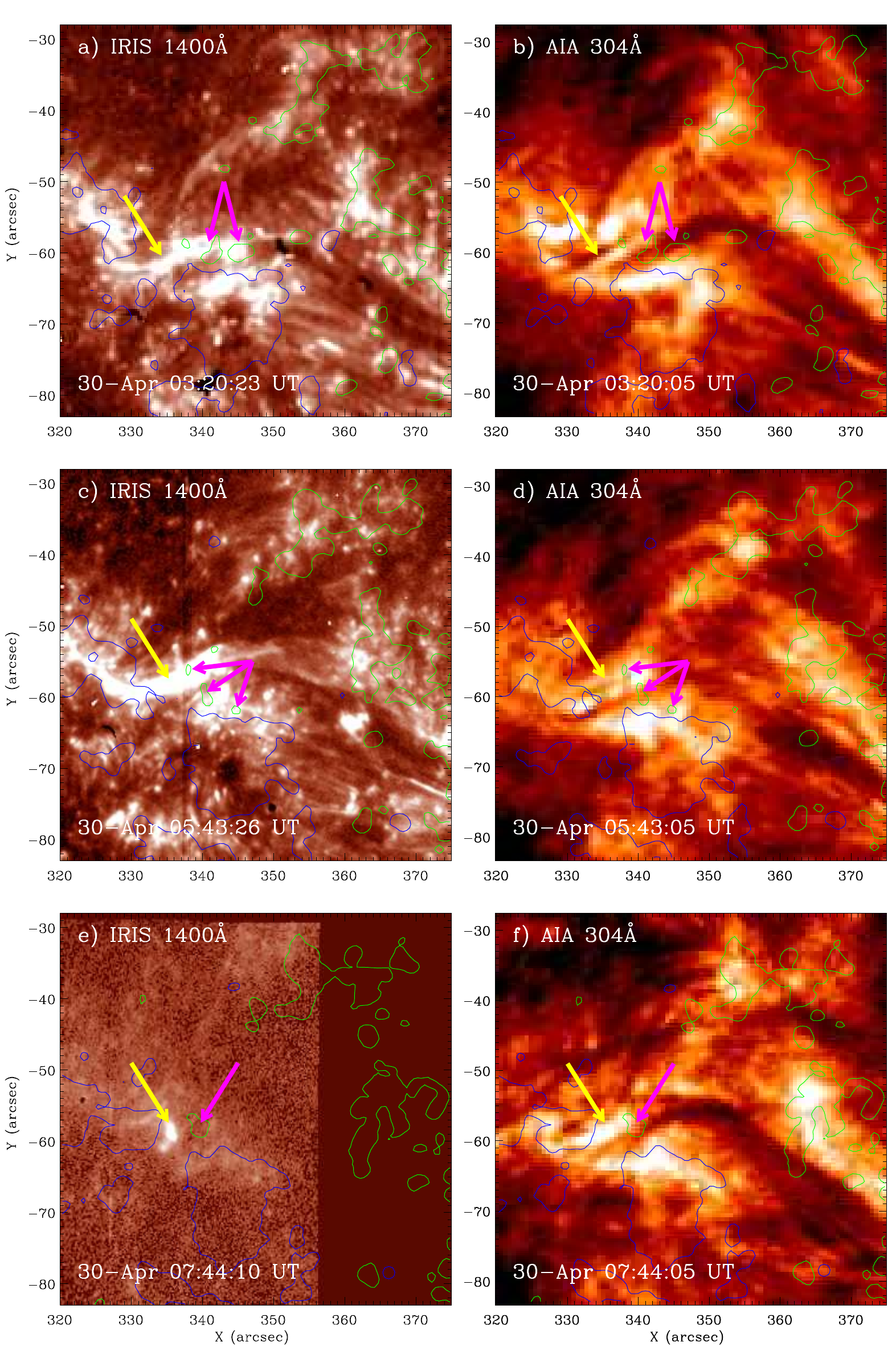}
    \caption{The upper-atmosphere response in AIA 304{\AA} (\emph{right panels}) and IRIS 1400{\AA} (\emph{left panels}). Contours of HMI longitudinal magnetic-fields at the closest time are superposed on all panels with positive (negative) fields in \emph{green (blue)}, and the levels for positive (negative) fields are 50 gauss. The \emph{yellow arrows} indicate the brightening. The \emph{pink arrows} indicate the micro positive polarities around N1 and N2. An accompanying animation is available in the Electronic Supplementary Material. The \emph{left and right panel} of the animation show the evolution of the IRIS 1400{\AA} and AIA 304{\AA}, respectively. The animation begins at 02:46:05 UT and ends at 08:10:05 UT on 30 April. The cadence is 16 seconds.}
    \label{fig7}
\end{figure}

As a result of continuous magnetic-flux cancellation and emergence beneath the adjacent ends of filaments, some brightenings successively appeared in the upper atmosphere. Three obvious brightenings before 08:00 UT on 30 April are exhibited in IRIS 1400{\AA} and AIA 304{\AA} (Figure~\ref{fig7} and Electronic Supplementary Material Animation 2). The contours of HMI magnetograms are superposed and the brightenings (yellow arrows) located between the micro positive polarities (pink arrows) and N1 and N2. The close spatial relationship likely demonstrates that the brightenings resulted from the magnetic-flux cancellation beneath. In addition, the brightenings in IRIS 1400{\AA} and AIA 304{\AA} were rarely seen in AIA 171{\AA}, which indicates that the brightenings plausibly occurred in the transition region.

\begin{figure}
	\includegraphics[width=\columnwidth]{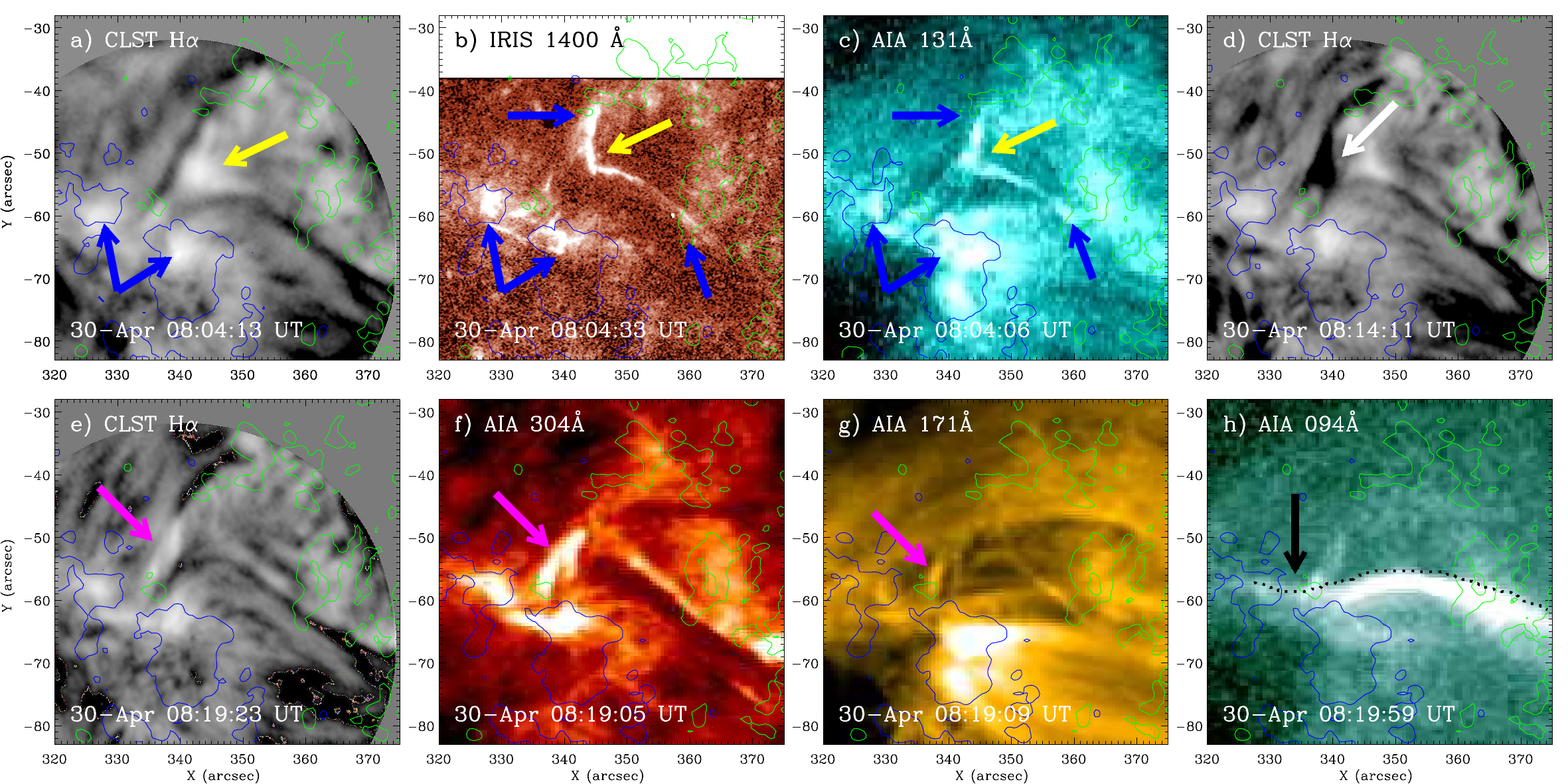}
    \caption{The interaction of two filaments in CLST H$\alpha$ filtergram \emph{\textbf{a, d, e}}, IRIS 1400{\AA} \emph{\textbf{b}}, and AIA 131, 304, 171, 94{\AA} \emph{\textbf{c}}, \emph{\textbf{f\,--\,h}}. The \emph{yellow arrows} indicate the brightening between two filaments. The \emph{blue arrows} indicate the brightening around the four ends of the two filaments. The \emph{white arrow} indicates the bridge of cold filament material connecting F1 and F2. The \emph{pink arrows} indicate the bright threads. The \emph{black arrow and dotted line} indicate the faint loops. Contours of HMI longitudinal magnetic fields at the closest time are superposed on all panels with positive (negative) fields in \emph{green (blue)}, and the levels for positive (negative) fields are 50 gauss. An accompanying animation is available in the Electronic Supplementary Material. The \emph{top panels} of the animation show the evolution of the CLST H$\alpha$ filtergram, IRIS 1400{\AA} and AIA 304{\AA}, respectively. The \emph{bottom panels} of the animation show the evolution of the AIA 171, 131, and 94{\AA}, respectively. The animation begins at 07:52:05 UT and ends at 08:30:05 UT on 30 April. The cadence is 6 seconds.}
    \label{fig8}
\end{figure}

Following the intermittent brightenings beneath the filament ends, there occurred an apparent interaction between two filaments (Figure~\ref{fig8} and Electronic Supplementary Material Animation 3). At $\approx$08:04 UT (Panels a\,--\,c), a bright point (yellow arrows) appeared from the low chromosphere to the high corona between F1 and F2, and linked with brightenings (blue arrows) around the four ends of the two filaments. It likely represents the X-type structure of the magnetic reconnection that occurred in the interface of the two filaments. Interestingly, some minutes later (Panel d), a bridge of cold filament material appeared connecting F1 and F2 (white arrow). The bridge quickly disappeared and was replaced by bright threads (pink arrows) that connected with the brightenings at the east end of F1 and the northern part of F2 (Panels f\,--\,g). Besides, some faint loops (black arrow and dotted line) appeared and connected the west end of F1 and the south end of F2 (Panel h). The newly formed bright threads and faint loops were likely the product of magnetic reconnection.

\begin{figure}
	\includegraphics[width=\columnwidth]{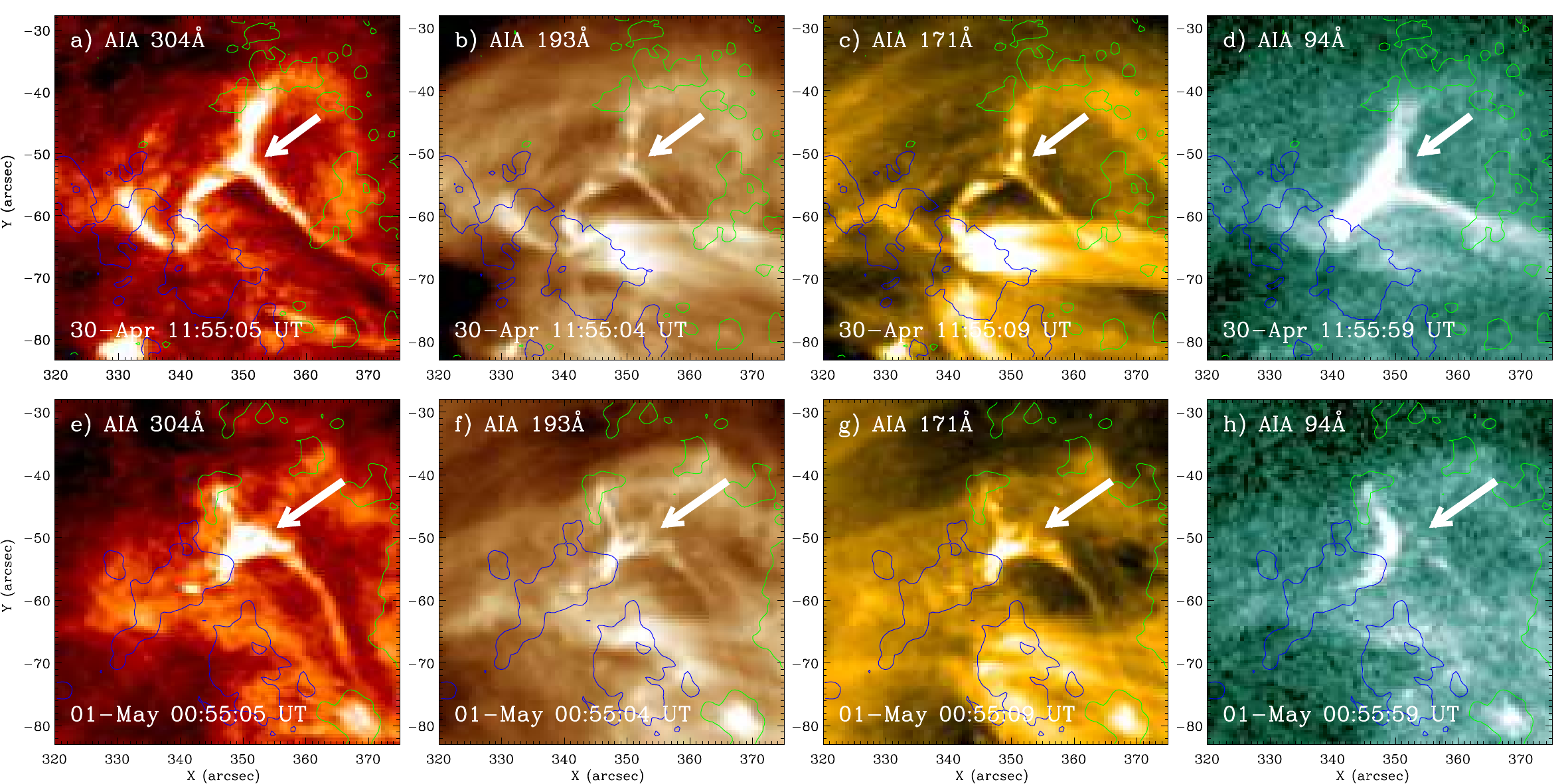}
    \caption{The intermittent interaction between two filaments in AIA 304, 171, 193, 94{\AA}. The \emph{white arrows} indicate the X-type brightenings. Contours of HMI longitudinal magnetic fields at the closest time are superposed on all panels with positive (negative) fields in \emph{green (blue)}, and the levels for positive (negative) fields are 50 gauss. An accompanying animation is available in the Electronic Supplementary Material. The \emph{four panels} of the animation show the evolution of AIA 304, 193, 171, and 94{\AA}. The animation consists of two time periods that begins at 11:40:04 UT and ends at 12:10:04 UT on 30 April, and begins at 00:40:04 UT and ends at 01:10:04 UT on 1 May. The cadence is 5 seconds.}
    \label{fig9}
\end{figure}

The interactions between two filaments occurred intermittently at some times in the following 20 hours, and two examples are shown in AIA 304, 171, 193, 94{\AA} (Figure~\ref{fig9} and Electronic Supplementary Material Animation 4). In each example, the X-type structure (white arrows) and brightenings around filament ends are clear evidence of magnetic reconnection between the two filaments.

\begin{figure}
	\includegraphics[width=\columnwidth]{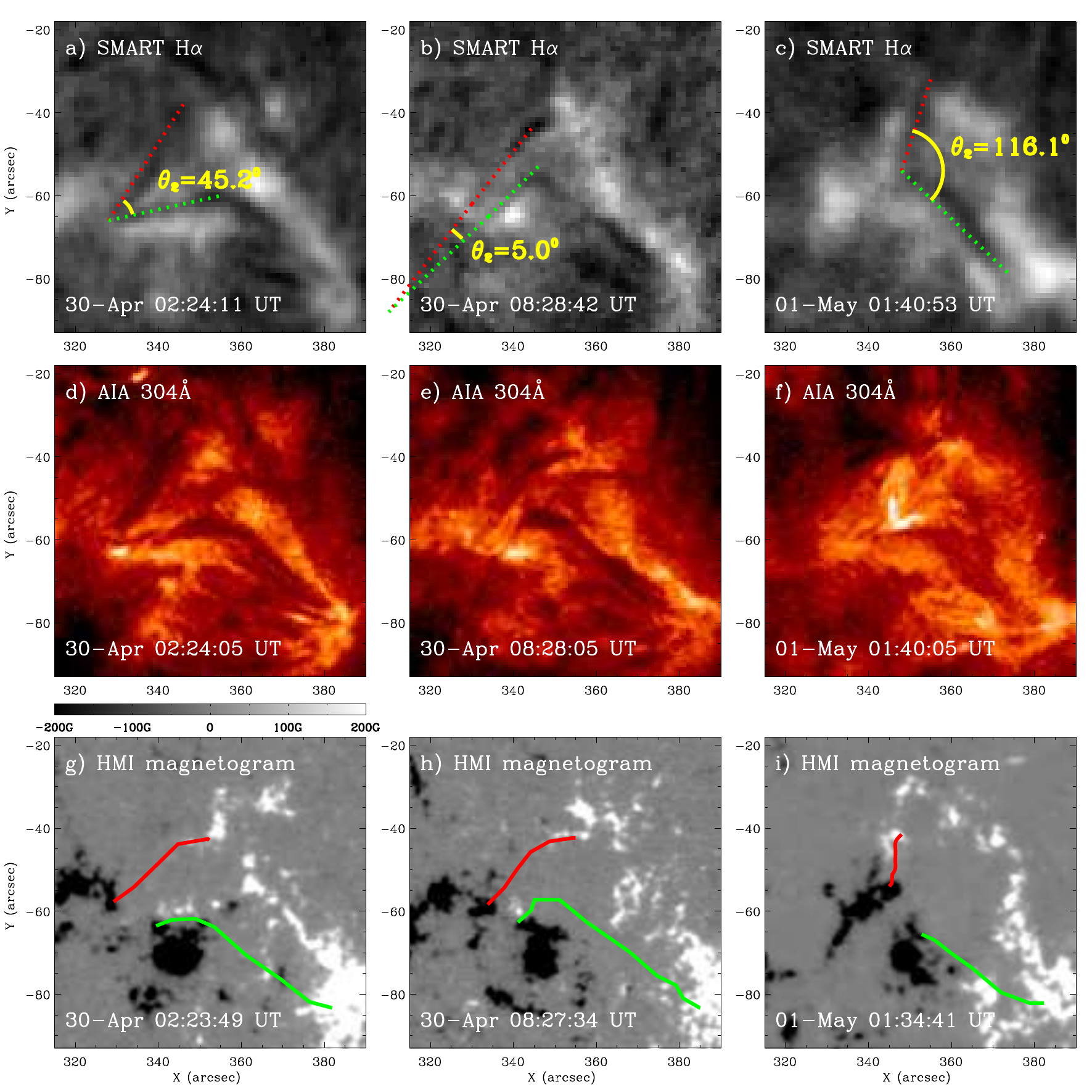}
    \caption{The overview of the morphology evolution of two filaments through a whole day in SMART H$\alpha$ filtergram (\emph{top panels}), AIA 304{\AA} (\emph{middle panels}), and HMI magnetogram maps (\emph{bottom panels}). The \emph{red and green-dotted lines} indicate the axis of F2 and the axis of eastern past of F1, respectively. \emph{$\theta_2$} indicates the angle between red and green-dotted lines. The \emph{green and red lines} indicate the profile of F1 and F2, respectively.}
    \label{fig10}
\end{figure}

Accompanying these successive interactions, two mini-filaments exhibited significant changes in their form. The morphology evolution of the two small filaments through a whole day is shown in Figure~\ref{fig10}. The changes of filaments are very clear in SMART H$\alpha$ and AIA 304{\AA} (top and middle panels), and the filament profiles (green and red lines) are superposed on HMI magnetograms (bottom panels). Remarkably, F1 was initially very sheared in its eastern part, and finally lessened in the shear during the evolution, and F2 rotated anti-clockwise and became short. In addition, the angle ($\theta_2$) between the F1 eastern part (green-dotted lines) and F2 (red-dotted lines) experienced two stages. Before the first interaction of filaments, $\theta_2$ decreased from $\approx$45.3\degree to $\approx$5.0\degree (Panels a\,--\,b). After the interactions, $\theta_2$ is roughly equal to the angle between F1 (the green-dotted line) and F2, which increased to $\approx$116.1\degree (Panel c). It is obvious that both F1 and F2 deformed after the successive interactions.
\section{Discussion and Conclusions}

In this article, we report the formation and immediate deformation of a small filament in AR 12760 on 28\,--\,30 April 2020. The filament formation in $\approx$40 hours was closely associated with the intermittent magnetic interactions between three emerging dipoles and the pre-existing magnetic polarities. The emergence of three dipoles added new magnetic energy for F1, and the separation motion drove and guided the coalescing process of the basic threads (FT0, FT1, FT2 and FT3). The cancellation between the emerging negative polarities of dipoles and previously-existing positive polarities led to the formation of the elongated F1. Such a coalescence process is very like the `head-to-tail' flux linkage scenario proposed by ~\cite{2001Martens}. Similar observations of the coalescence formation process of large and small filaments can be seen in the work of~\cite{2007wang} and~\cite{2020chena}.

The newly-formed F1 quickly deformed the following day.
The continuous magnetic-flux cancellation beneath the adjacent ends of the F1 and the nearby F2 likely destabilized the F1 and F2 and thus intermittent magnetic interactions occurred between them. Finally, F1 lessened in the shear, and F2 became much shorter (Figure 10). Hence, the filament F1 experienced the formation and immediate deformation through intermittent magnetic interactions.

Based on the observational results and discussion, we propose a possible scenario of the formation of F1 with three main steps (Figure 11). In the first step (upper panels), there existed some filament threads (FT0) connecting P0a and N0a, before the emergence of the filament threads (FT1) of the first dipole (P1a-N1a). The eastward N1a interacted with P0a (the yellow star), and magnetic reconnection occurred between FT0 and FT1. As a result, the filament in the first phase (F1$^{\prime}$) formed and connected N0a and P1a, with a dip above the interaction location between P0a and N1a. In the second step (middle panels), N1a kept interacting with P0a and P0b, and N0a also interacted with P0a and P0b (yellow stars). Besides, N1b of the second dipole also moved eastwards and mixed with N1a, and P1b merged with P0c and P1a. Therefore, the eastern part of the filament in the second phase (F$^{\prime\prime}$) became convex. In the third step (bottom panels), the emergence of the third dipole brought in the filament threads (FT3), and eastward movement of N1c led to the interaction between N1c and P0b and the magnetic reconnection between FT3 and F$^{\prime\prime}$ (the yellow star). Accompanying with the convergence, an elongated filament (F1) finally formed through intermittent magnetic interactions associated with the successive emergence of three dipoles. In this discussion we simply focus on the main footpoints of the filaments. As discussed by~\cite{2020chenb}, active-region filaments generally have few or no other footpoints (i.e. barbs), and the filaments we discuss here fits that pattern.

The deformation process of F1 is shown in Figure 12. The eastern part of F1 was very sheared and close to the southern part of F2. Due to the continuous magnetic-flux cancellation beneath the adjacent ends of two filaments and the convergence (black arrows) of the two ends of F2, magnetic interactions (yellow star) intermittently occurred between the eastern part of F1 and F2 (Panel a). As the production of magnetic reconnections, a longer loop (the red line) connected the western end of F1 and the southern end of F2, and a short loop (the blue-dotted line) linked the eastern end of F1 and the northern end of F2, which is consistent with the newly formed occasional structures (pink and black arrows in Figure 8). Finally, F1 lessened in the shear, and F2 became shorter (Panel b).

Obviously, there was a close relationship between the formation and deformation of F1 and the continuous magnetic activities including the emergence, cancellation, and convergence. However, the same types of magnetic activities played different roles in two stages of filament evolution. They led to a head-to-tail reconnection of a pair of dipoles and are responsible for the development of the elongated F1. It was a process that slowly accumulated magnetic energy in the filament structure and made the filament stable, but it slowly disturbed the two filaments in the deformation process and gradually released magnetic energy by intermittent magnetic reconnections.

Moreover, it is generally believed that the eruption involved an accumulation process of magnetic energy. When the stored magnetic energy reaches a certain stage, loss of equilibrium occurs in the system accompanying with magnetic reconnections, and the magnetic energy can be intensely released in the form of an eruption~\citep{2011chen}. However, the energy accumulation process may not always smoothly. Like the F1 in this case, it only experienced deformation, not eruption, after intermittent magnetic reconnections. Hence, it is possible that the stored energy in the filament might be dissipated by small-scale reconnection in some cases and thus forestall the eruption. It is important to note this possibility.

In summary, we believe that the formation and immediate deformation of a small filament was closely associated with its magnetic-flux emergence, cancellation, and convergence. We also suggest that the gradual release of magnetic energy is an important factor if a filament is to only deform but not erupt. The high resolution 1.8-meter C\emph{hinese} L\emph{arge} S\emph{olar} T\emph{elescope} (CLST) captured only one of the interaction processes, therefore continuous high-resolution observations of deformation process still needs to be studied. Other numerical methods such as MHD simulation, or nonlinear force-free field (NLFFF) methods should be employed to measure the energy loss in the deformation process.
More studies are needed to understand the process of formation and deformation of filaments.
\begin{acks}
Many thanks to Shuhong Yang and Xiaoshuai Zhu for a constructive discussion. We gratefully acknowledge the usage of data from the SDO and IRIS spacecraft, and the ground-based CLST and SMART projects. This work is supported by grants NSFC 11790303, 11727305 and 12073016.
\end{acks}

\begin{ethics}
\begin{conflict}
The authors declare that they have no conflicts of interest.
\end{conflict}
\end{ethics}

\begin{dataavailability}
The datasets generated during and/or analysed during the current study are available in the [JSOC, Official website of IRIS and SMART] repository, [\url{jsoc.stanford.edu}, \url{iris.lmsal.com}, \url{www.hida.kyoto-u.ac.jp/SMART}].
The CLST data are available from the author on request by email.
\end{dataavailability}

\bibliographystyle{spr-mp-sola}

\clearpage
\clearpage

 \clearpage

\clearpage
\clearpage
\clearpage
\clearpage

\clearpage

\clearpage

\clearpage

\clearpage


\begin{figure}
	\includegraphics[width=\columnwidth]{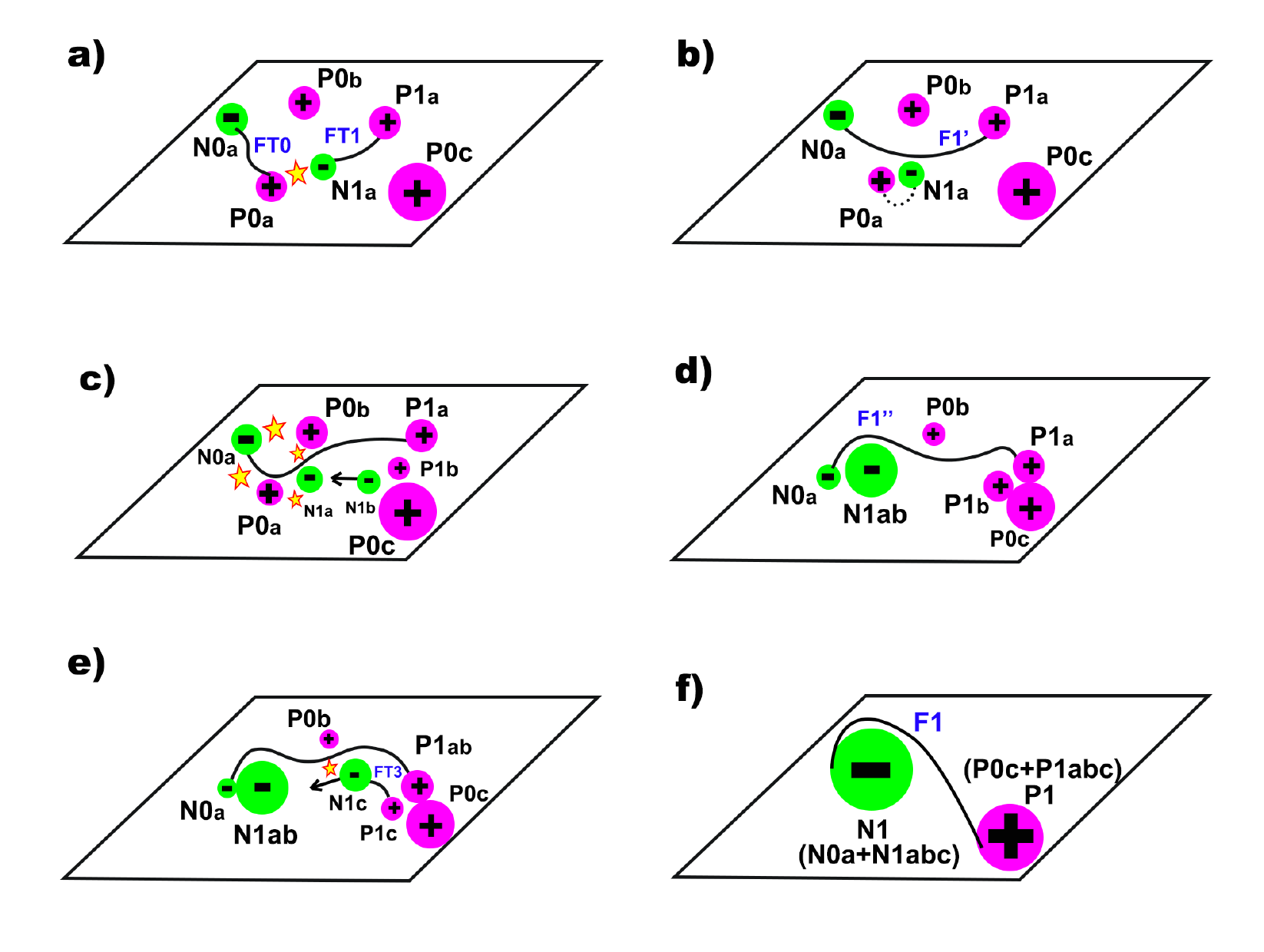}
    \caption{Schematic showing the scenario of the formation of filament F1 associated with the emergence of three dipoles (N1a-P1a, N1b-P1b, N1c-P1c), their quick separation, and the cancellation of the emerged negative polarities and pre-existing positive polarities (P0a, P0b). The \emph{green and pink balls and the symbols} above them indicate the positive and negative polarities, respectively. The \emph{yellow stars} indicate the magnetic cancellation sites. The \emph{black lines} indicate the filament. The \emph{black arrows} indicate the direction of motion of the polarities.}
    \label{fig11}
\end{figure}

\clearpage


\begin{figure}
	\includegraphics[width=\columnwidth]{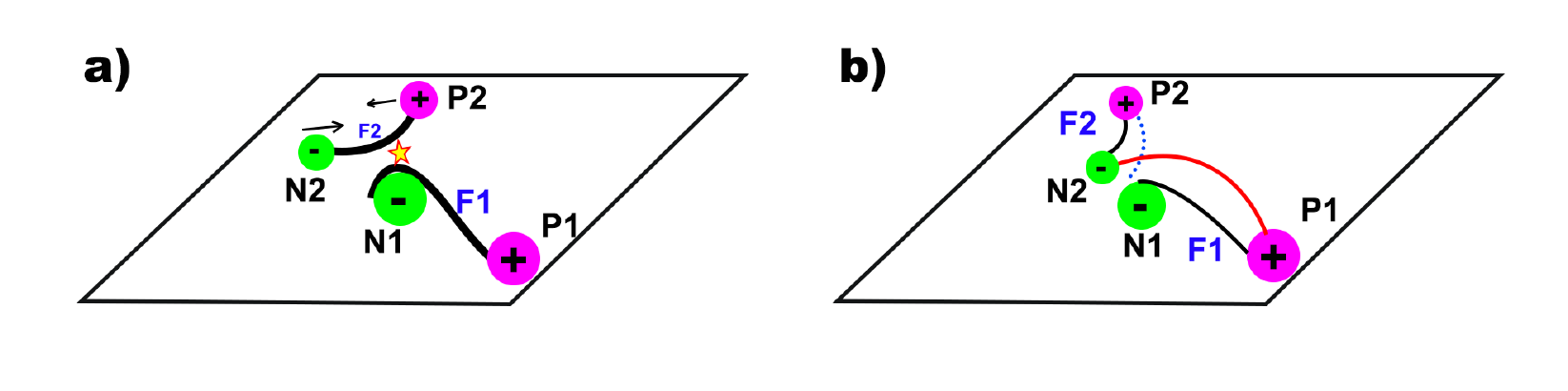}
    \caption{Schematic showing the scenario of the deformation of filament F1 associated with the continuous magnetic-flux cancellation beneath the adjacent ends of two filaments and the convergence of two ends of F2. The \emph{green and pink balls and the symbols} above them indicate the positive and negative polarities, respectively. The \emph{black arrows} indicate convergence motion. The \emph{black lines} indicate the filament. The \emph{yellow stars} indicate the magnetic-reconnection sites. The \emph{red line and blue-dotted line} indicate two sets of loops that are the product of magnetic reconnection.}
    \label{fig12}
\end{figure}
\clearpage

\end{article}

\end{document}